\documentclass[aps,prl,twocolumn,superscriptaddress]{revtex4}

\usepackage{graphicx}
\usepackage{epsfig}
\usepackage{here}

\begin{document}

\title{Measurement of $R = \sigma_L / \sigma_T$  in
                Deep-Inelastic Scattering on Nuclei
      }


\def\groupalberta{\affiliation{Department of Physics, University of Alberta, Edmonton, Alberta T6G 2J1, Canada}}
\def\groupargonne{\affiliation{Physics Division, Argonne National Laboratory, Argonne, Illinois 60439-4843, USA}}
\def\groupbari{\affiliation{Istituto Nazionale di Fisica Nucleare, Sezione di Bari, 70124 Bari, Italy}}
\def\groupcolorado{\affiliation{Nuclear Physics Laboratory, University of Colorado, Boulder, Colorado 80309-0446, USA}}
\def\groupdesy{\affiliation{DESY, Deutsches Elektronen-Synchrotron, 22603 Hamburg, Germany}}
\def\groupzeuthen{\affiliation{DESY Zeuthen, 15738 Zeuthen, Germany}}
\def\groupdubna{\affiliation{Joint Institute for Nuclear Research, 141980 Dubna, Russia}}
\def\grouperlangen{\affiliation{Physikalisches Institut, Universit\"at Erlangen-N\"urnberg, 91058 Erlangen, Germany}}
\def\groupferrara{\affiliation{Istituto Nazionale di Fisica Nucleare, Sezione di Ferrara and Dipartimento di Fisica, Universit\`a di Ferrara, 44100 Ferrara, Italy}}
\def\groupfrascati{\affiliation{Istituto Nazionale di Fisica Nucleare, Laboratori Nazionali di Frascati, 00044 Frascati, Italy}}
\def\groupfreiburg{\affiliation{Fakult\"at f\"ur Physik, Universit\"at Freiburg, 79104 Freiburg, Germany}}
\def\groupgent{\affiliation{Department of Subatomic and Radiation Physics, University of Gent, 9000 Gent, Belgium}}
\def\groupgiessen{\affiliation{Physikalisches Institut, Universit\"at Gie{\ss}en, 35392 Gie{\ss}en, Germany}}
\def\groupglasgow{\affiliation{Department of Physics and Astronomy, University of Glasgow, Glasgow G128 QQ, United Kingdom}}
\def\groupillinois{\affiliation{Department of Physics, University of Illinois, Urbana, Illinois 61801, USA}}
\def\groupliverpool{\affiliation{Physics Department, University of Liverpool, Liverpool L69 7ZE, United Kingdom}}
\def\groupwisconsin{\affiliation{Department of Physics, University of Wisconsin-Madison, Madison, Wisconsin 53706, USA}}
\def\groupmit{\affiliation{Laboratory for Nuclear Science, Massachusetts Institute of Technology, Cambridge, Massachusetts 02139, USA}}
\def\groupmichigan{\affiliation{Randall Laboratory of Physics, University of Michigan, Ann Arbor, Michigan 48109-1120, USA }}
\def\groupmoscow{\affiliation{Lebedev Physical Institute, 117924 Moscow, Russia}}
\def\groupmunich{\affiliation{Sektion Physik, Universit\"at M\"unchen, 85748 Garching, Germany}}
\def\groupnikhef{\affiliation{Nationaal Instituut voor Kernfysica en Hoge-Energiefysica (NIKHEF), 1009 DB Amsterdam, The Netherlands}}
\def\groupstpetersburg{\affiliation{Petersburg Nuclear Physics Institute, St. Petersburg, Gatchina, 188350 Russia}}
\def\groupprotvino{\affiliation{Institute for High Energy Physics, Protvino, Moscow oblast, 142284 Russia}}
\def\groupregensburg{\affiliation{Institut f\"ur Theoretische Physik, Universit\"at Regensburg, 93040 Regensburg, Germany}}
\def\grouprome{\affiliation{Istituto Nazionale di Fisica Nucleare, Sezione Roma 1, Gruppo Sanit\`a and Physics Laboratory, Istituto Superiore di Sanit\`a, 00161 Roma, Italy}}
\def\groupsimonfraser{\affiliation{Department of Physics, Simon Fraser University, Burnaby, British Columbia V5A 1S6, Canada}}
\def\grouptriumf{\affiliation{TRIUMF, Vancouver, British Columbia V6T 2A3, Canada}}
\def\grouptokyo{\affiliation{Department of Physics, Tokyo Institute of Technology, Tokyo 152, Japan}}
\def\groupamsterdam{\affiliation{Department of Physics and Astronomy, Vrije Universiteit, 1081 HV Amsterdam, The Netherlands}}
\def\groupyerevan{\affiliation{Yerevan Physics Institute, 375036 Yerevan, Armenia}}


\groupalberta
\groupargonne
\groupbari
\groupcolorado
\groupdesy
\groupzeuthen
\groupdubna
\grouperlangen
\groupferrara
\groupfrascati
\groupfreiburg
\groupgent
\groupgiessen
\groupglasgow
\groupillinois
\groupliverpool
\groupwisconsin
\groupmit
\groupmichigan
\groupmoscow
\groupmunich
\groupnikhef
\groupstpetersburg
\groupprotvino
\groupregensburg
\grouprome
\groupsimonfraser
\grouptriumf
\grouptokyo
\groupamsterdam
\groupyerevan


\author{A.~Airapetian}  \groupyerevan
\author{N.~Akopov}  \groupyerevan
\author{Z.~Akopov}  \groupyerevan
\author{M.~Amarian}  \grouprome \groupyerevan
\author{V.V.~Ammosov}  \groupprotvino
\author{E.C.~Aschenauer}  \groupzeuthen
\author{R.~Avakian}  \groupyerevan
\author{A.~Avetissian}  \groupyerevan
\author{E.~Avetissian}  \groupyerevan
\author{P.~Bailey}  \groupillinois
% ...

\collaboration{The HERMES Collaboration} \noaffiliation

\date{\today}

\begin{abstract}
Cross section ratios for deep-inelastic scattering from
$^3$He, $^{14}$N and $^{84}$Kr with respect to $^2$H have been measured
by the HERMES collaboration at DESY using a 27.5 GeV positron beam.
The data cover a range in the Bjorken scaling
variable $x$ between 0.010 and 0.65, the negative squared
four-momentum transfer $Q^2$ varies from 0.5 to 15 GeV$^2$,
while at small values of $x$ and $Q^2$, 
the virtual photon polarisation parameter $\epsilon$ extends
to lower values than previous measurements.
From the dependence of the data on
$\epsilon$,
values for $R_A/R_D$ with $R$ the ratio $\sigma_L / \sigma_T$ of longitudinal
to transverse DIS cross sections have been derived
and found to be consistent with unity.
\end{abstract}


\maketitle

\newpage
In 1993 the EMC collaboration published a significant difference between the 
deep-inelastic scattering (DIS) cross sections per nucleon on iron and 
deuterium~\cite{emc-effect}, 
indicating that the quark momentum distributions in bound nucleons differ 
from those in free nucleons. 
This phenomenon is known as the {\it EMC effect} at large values of the
Bjorken scaling variable $x (x > 0.1)$, and as {\it shadowing}
at lower values of $x$ \cite{Arne94}.


With $F_2(x)$ found to be $A$-dependent, it is relevant to investigate
whether this dependence is the same for its longitudinal 
and transverse components, $F_L(x)$ and $F_1(x)$.
The latter two structure functions are related to $F_2(x)$ via
$F_L(x) = (1 + Q^2/\nu^2) F_2(x) - 2 x F_1(x)$
with $-Q^2$ the square of the four-momentum transfer, 
$\nu$ the energy transfer, $x = Q^2 / 2 M \nu$ and $M$ the nucleon
mass. A possible difference between the $A$-dependences of 
$F_L(x)$ and $F_1(x)$ can be investigated by measuring
the ratio of longitudinal to transverse deep-inelastic
scattering cross sections 
$R = \sigma_L / \sigma_T = F_L(x) / 2 x F_1(x)$
for various nuclear targets. 
While several such studies have been reported \cite{SLAC,AuD,SnC}, 
none of these
has provided values of $R$ in the kinematic regime $x<0.06$ with
$Q^2<1$\,GeV$^2$. 
This region is the focus of the present paper.

In deep-inelastic
charged lepton scattering from an unpolarised target,
the double-differential cross section per nucleon can be written 
in the one-photon exchange approximation as
\begin{eqnarray}
    \frac{d^2\sigma}{dxdQ^2} & = &
    \frac{4\pi\alpha^2}{Q^4} \frac{ F_2(x,Q^2)} {x} \times \nonumber\\
    & & \left[
            1-y- \frac{Q^2}{4E^2}  + 
                  \frac{y^2}{2} \left(
            \frac{   1+ Q^2 / \nu^2 } {1+R(x,Q^2)} \right)
         \right] \nonumber\\
    & = & \frac{\sigma_{\rm{Mott}}}{E^{\prime} E} 
	  \frac{\pi F_2(x,Q^2)}{x \epsilon}
	   \frac{1 + \epsilon R(x,Q^2)}{1 + R(x,Q^2)} ,
\end{eqnarray}
where $y = \nu / E$, $\sigma_{\rm{Mott}}$ represents the cross section for
lepton scattering from a point charge,
and $E$ and $E^{\prime}$ are the initial and final lepton energies
in the target rest frame, respectively. 
The virtual photon polarisation 
parameter is given by
\begin{equation}
	\epsilon = \frac{ 4(1-y) - Q^2/E^2 }
			{4(1-y) + 2y^2 + Q^2/E^2 }.
\end{equation} 
The ratio of DIS cross sections on nucleus $A$ and deuterium $D$ (=$^2$H)
is then given by:
\begin{equation}
\frac{\sigma_{A}}{\sigma_{D}}=\frac{F_2^A}{F_2^D}
        \frac{(1+ \epsilon R_A)(1+R_D)}{(1+R_A)(1+\epsilon R_D)},
\end{equation}
where $R_A$ and $R_D$ represent the ratio $\sigma_L / \sigma_T$ for
nucleus $A$ and deuterium.
To facilitate easier interpretation and in accordance to the literature, 
here and throughout this paper all cross sections 
are defined as cross sections per nucleon and are 
converted to cross sections for isoscalar nuclei, i.e. the measured cross 
sections are  
divided by the atomic number $A$ and corrected for any difference in the 
number of protons and neutrons:
\begin{equation} 
\frac{\sigma_{A}}{\sigma_{D}} \equiv \frac{\sigma_{A}^{nucleus}}
                       {Z \sigma_{p} + (A-Z) \sigma_{n}} , 
\end{equation}
where $\sigma_{A}^{nucleus}$ is the DIS cross section per nucleus for nucleus A and  
$\sigma_{p}$ and $\sigma_{n}$ are the DIS cross sections on the proton and the neutron. 
In practice, $\sigma_{A}^{nucleus}/\sigma_{D}$ is converted to 
$\sigma_{A}/\sigma_{D}$ using the 
known cross section ratio 
$\sigma^D/\sigma^p$ \cite{nmcf2np}. 

For $\epsilon \rightarrow$ 1 the cross section ratio equals the
ratio of structure functions $F_2^A / F_2^D$.
For smaller values
of $\epsilon$ the cross section ratio is equal 
to $F_2^A / F_2^D$ only if $R_A = R_D$. 
A difference between $R_A$ and $R_D$ will thus
introduce an $\epsilon$-dependence of $\sigma_{A}/\sigma_{D}$.
Hence, measurements of $\sigma_{A}/\sigma_{D}$ as a function of
$\epsilon$ can be used to extract experimental information 
on $R_A / R_D$, if $R_D$ is known.

In this paper we present data from the HERMES experiment
on the cross section ratio for
deep-inelastic positron scattering off helium-3, nitrogen and krypton 
with respect to deuterium. 
The helium-3 and nitrogen data were published in  
a previous letter \cite{oldpaper}. Recently, those data were found to be subject 
to an A-dependent tracking inefficiency of the HERMES 
spectrometer \cite{erratum}, which was not recognised in the previous 
analysis. The resulting correction of the cross section 
ratios is significant at low values of $x$ and $Q^2$ and 
substantially changes the interpretation of those data.
The data presented here were corrected for this effect and 
supersede those published in Ref. \cite{oldpaper}. 

The data 
were collected by the HERMES experiment
at DESY using $^1$H, $^2$H, $^3$He, 
$^{14}$N and $^{84}$Kr molecular gas targets internal to the 27.5 GeV 
positron storage ring of HERA. 
The target gases were injected
into a 40 cm long, tubular open-ended
storage cell inside the positron ring.
The luminosity was measured by detecting Bhabha-scattered
target electrons in coincidence with the scattered positrons, in a pair of 
NaBi(WO$_4$)$_2$ electromagnetic calorimeters. 

The HERMES spectrometer \cite{specpaper} is a forward angle instrument 
which is symmetric about a central horizontal shielding plate in the 
magnet. 
Both the scattered positrons 
and the hadrons produced can be detected and identified within an
angular acceptance of $\pm$ 170 mrad horizontally, and 40 -- 140 mrad
vertically.
The trigger was formed from a coincidence between signals of three scintillator
hodoscope planes and a lead-glass calorimeter
where a minimum energy deposit 
of 3.5 GeV was required.
Positron identification was accomplished using 
the calorimeter, the second hodoscope, which functioned as a 
preshower counter, a transition-radiation detector
and a \u Cerenkov counter. This system provided
positron identification with an average efficiency of 99~\% and a
hadron contamination of less than 1~\%.

Deep-inelastic scattering events were
extracted from the data by imposing constraints on
$Q^2$ $(Q^2 > 0.3$ GeV$^2$), 
$W$ (the invariant mass of the photon-nucleon system, required to be greater 
than 2 GeV),
and $y$ $(y<0.85)$. 
At very low $x$ and high $y$, the nuclear cross sections are dominated by 
radiative processes associated with elastic scattering. 
To limit these contributions, a minimum $x$ value of $x=0.01$ was 
required.

As the ratio $\sigma_A / \sigma_D$ involves nuclei with different
numbers of protons, radiative corrections do not cancel in the
ratio. In particular, the yield of radiative processes associated with
elastic scattering scales with $Z^2$ and thus differs for the two
target nuclei. 
At small values of apparent $x$ and $Q^2$ (inferred from the kinematics of the scattered 
positron),
corresponding to large values of $y$,
the contribution from radiative elastic scattering becomes large.
In this kinematic region, the associated energetic photons
radiated at small angles can produce electromagnetic showers
that cause large tracking inefficiencies~\cite{erratum}.
Corrections for these process-specific inefficiencies must be
applied, since they increase in severity as the $Z$ of nuclear
targets increases.

\begin{figure}[H]
\begin{center}
\includegraphics[width=0.47\textwidth,height=8.0cm]{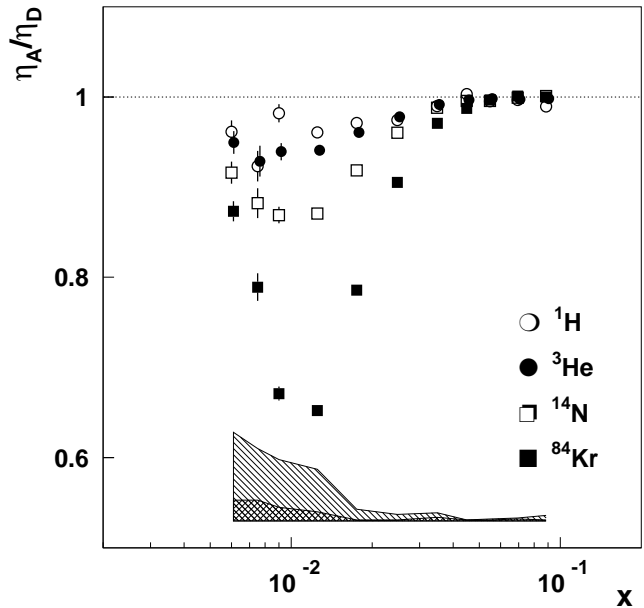}
\epsfxsize=\hsize
\end{center}
\vspace*{-0.2cm}
\caption{
Ratio of track reconstruction efficiencies in $^1$H, $^3$He, $^{14}$N and $^{84}$Kr  
with respect to $^2$H as function of $x$.
The hatched areas represent the systematic uncertainties for the N/D 
(cross hatched) 
and Kr/D (slanted hatched) efficiency 
ratios. The systematic uncertainties for the H/D and He/D 
ratios are smaller than that of the N/D ratio and are not shown.}
\label{fig:losses}
\end{figure}

These track reconstruction losses in the HERMES detector have been simulated 
in detail using the GEANT-based Monte Carlo description of the 
experiment. The probability of photon emission is modelled following 
the description of Mo and Tsai \cite{MoTsai}, and has been carefully 
compared to other calculations of radiative processes. 
All materials close to the beam pipe have been implemented in detail
and the effect of the 
minimum energy of the secondary particles tracked through the 
detector was investigated. 
The resulting reconstruction losses at low $x$ and $Q^2$ strongly 
depend on the target material and show a strong variation with $y$, 
and consequently with $x$ and $Q^2$.
The ratios of the reconstruction efficiencies $\eta$ for target nucleus A compared 
to deuterium are shown 
in Fig.~\ref{fig:losses} as a function of $x$, 
for the various target materials used in the HERMES experiment. 
For completeness, this figure includes points at smaller
values of $x$ than are employed in the present analysis.

\begin{figure}[htb]
\begin{center}
\includegraphics[width=0.47\textwidth,height=8.0cm]{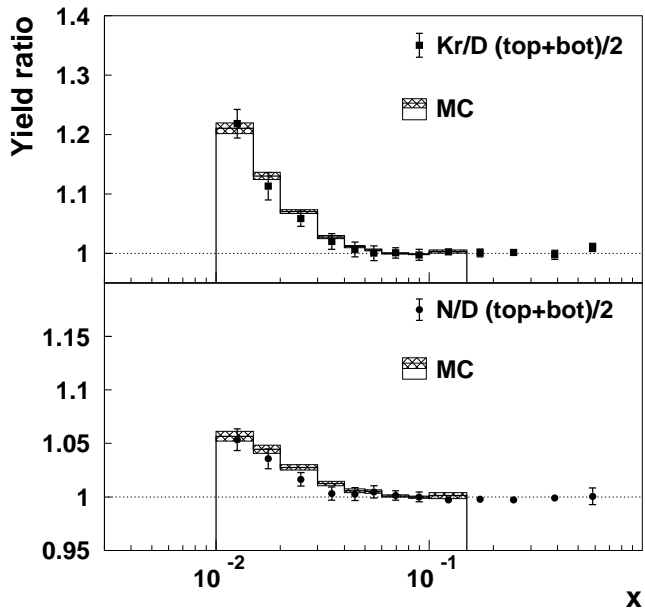}
\epsfxsize=\hsize
\end{center}
\vspace*{-0.2cm}
\caption{
Comparison between data (points) and MC simulation (histogram) for the 
fractional change in the cross section ratios when treating the 
upper and lower HERMES detector halves independently.}
\label{fig:superratio}
\end{figure}

The systematic uncertainty of this correction was estimated using the 
fact that the HERMES spectrometer consists of  two
independent detectors above and below the positron beam.
For about 50~\% of the events with a hard radiated photon the resulting 
electromagnetic shower is contained in one detector while the scattered  
electron is found in the other detector. 
While these events are rejected by the standard HERMES reconstruction algorithm 
because of their high total multiplicity, they can be 
reconstructed when one considers the two detectors independently. 
The number of events gained in this way strongly depends on the 
details of the electromagnetic shower -- especially on the energy of the radiated photon and the exact position where the photon hits any material  --      
and thus provides 
a good measure of the quality of the MC simulation. 
Reasonable agreement between the data and the simulation is found for all
target materials.
Fig.~\ref{fig:superratio} 
shows as a function of apparent $x$ the fractional change in the yield ratios 
when treating the upper and lower spectrometer halves independently, 
both for the data and the MC simulation. 
Here the yields from the two detector halves have been averaged.
A difference between the yields in the upper and the lower detector 
observed in the data is attributed to a relative misalignment between 
the two detectors.
As is typical practice for combining results which are statistically 
incompatible~\cite{PDG}, the statistical uncertainty of the weighted mean 
is scaled by the square root of $\chi^2$. 
The difference between the yields in the two detectors is significant
only in the krypton data at low $x$ ($x < 0.03$) where the reconstruction
efficiency quickly changes with $x$ and the difference can be up
to 10~\%.
The difference between the data and the MC simulation is
treated as a systematic uncertainty. 

The track reconstruction inefficiency mainly affects radiative 
elastic and to some extent quasielastic events.  
These and all other radiative processes have been 
computed using the method outlined in  Ref. \cite{ABA}.
For the evaluation of the coherent radiative tails, 
the nuclear elastic form factors must be known. Parameterisations of 
the form factors of $^2$H, $^3$He, $^{14}$N and $^{84}$Kr were taken from the 
literature \cite{deut,heli,nitr,kryp}. 
For the quasi-elastic tails, the nucleon form factor parameterisation of 
Gari and Kr\"umpelmann \cite{gari} was used. The reduction of the 
bound nucleon cross section with respect to the free nucleon one 
(quasi-elastic suppression) was evaluated using the results of a calculation 
by Bernabeu \cite{berna} for deuterium and the 
non-relativistic Fermi gas model for $^3$He, $^{14}$N  and $^{84}$Kr \cite{moniz}.
The evaluation of the inelastic QED
processes requires the knowledge 
of both $F_2$ and $R$ over a wide range of $x$ and $Q^2$. 
The structure function $F_2^D(x,Q^2)$ was described by a 
Regge motivated parameterisation of the proton structure function 
$F_2^p(x,Q^2)$ \cite{f2allm}
multiplied by the NMC measurement of $F_2^D/F_2^p$ \cite{nmcf2np};
for $R_D$  the Whitlow 
parameterisation \cite{R1990} was used. 
The nuclear structure 
functions $F_2^A(x,Q^2)$ were taken from phenomenological fits to the 
SLAC and NMC data, and $R_A(x,Q^2)$ was assumed to be equal to $R_D(x,Q^2)$.
The effects of all radiative processes were subtracted from the measured 
yields and the statistical errors propagated accordingly. 
This method avoids the possible large model dependence that can result 
from multiplicatively applying radiative corrections \cite{andy}. 
Because of the reconstruction inefficiency explained above, only those 
radiative events actually seen by the HERMES spectrometer were 
subtracted. 

For high-$Z$ targets such as krypton, the probability for the exchange of more than 
one photon becomes non-negligible. 
The corresponding amplitudes lead to a
destructive interference with the leading amplitudes. 
For the dominant contribution to the radiative corrections --- the coherent 
radiative tail --- this effect has been estimated using the distorted 
wave function method \cite{kurek}, resulting into a 5-10~\% reduction of 
the radiative elastic tail. 
Other contributions proportional to $Z\cdot\alpha_{EM}$ might be non-negligible but 
could not be estimated. 

The systematic uncertainty in the radiative corrections was estimated
by using upper and lower limits of the parameterisations,
or alternative parameterisations \cite{NMCpar,R1990,diff,ArDay}
for all the above input parameters.
The resulting systematic uncertainty in the cross section ratio
of Kr/D is about   
6~\% at low $x$, quickly falling to values smaller than 1~\% for $x>0.06$.
(As mentioned before this uncertainty does not include the effects of 
multiphoton exchange (Coulomb distortion) beyond the estimated contribution  
to the coherent elastic tail; these contributions might be non-negligible but could 
not be estimated.)
Because of the smaller radiative contributions, the uncertainties due to the 
radiative corrections in 
the cross section 
ratios of nitrogen and helium with respect to deuterium are about 3~\% 
at low $x$. 

The effects originating from the finite resolution of 
the spectrometer and from the hadron contamination in the positron sample 
have been determined and found to be negligible.
The overall normalisation uncertainty has been estimated from the 
luminosity measurements to be 1.4~\% for the He(N)/D 
data and 1~\% for the Kr/D data.

\begin{figure} [H]
\begin{center}
\includegraphics[width=0.47\textwidth]{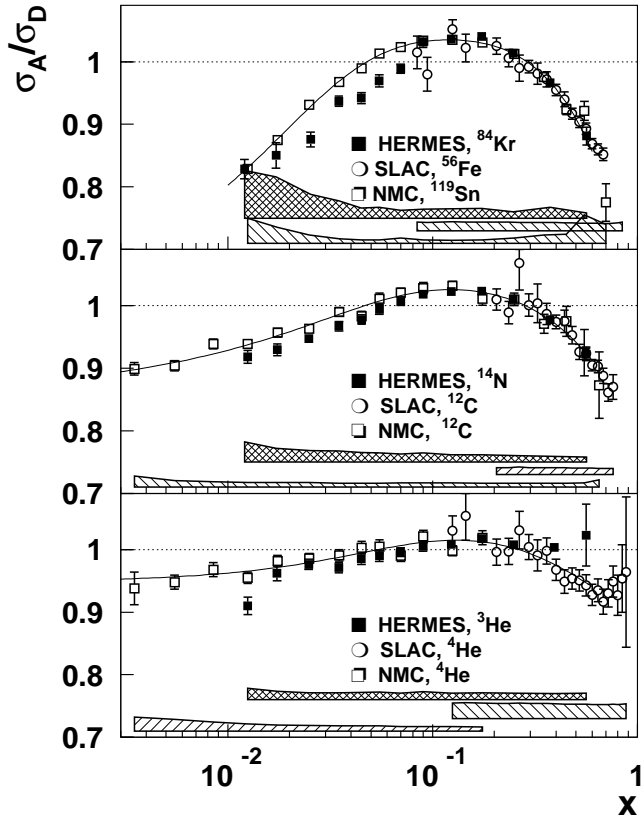}
\epsfxsize=\hsize
\end{center}
\vspace*{-0.2cm}
\caption{
Ratio of isoscalar Born cross sections of
inclusive deep-inelastic lepton scattering from nucleus
$A$ and $D$ versus $x$.
The error bars represent the statistical
uncertainties, the systematic uncertainties are given by the
error bands.  
The HERMES $^3$He/D and $^{14}$N/D data have been renormalised by
        0.9~\%.
}
\label{fig:xdep}
\end{figure}

The results of the present analysis \cite{erika} 
are shown in Fig.~\ref{fig:xdep} as a
function of $x$. 
Also shown are the results of the NMC \cite{nmche,nmccc,nmcsn} 
and SLAC \cite{slace139} measurements of 
$\sigma_{He}/\sigma_D$, 
$\sigma_C/\sigma_D$  
and $\sigma_{Sn}/\sigma_D$ where the NMC values for 
$\sigma_{Sn}/\sigma_D$ have been obtained from the measurements of 
$\sigma_{Sn}/\sigma_C$ and $\sigma_C/\sigma_D$.
On average, the present data on $\sigma_{He}/\sigma_D$ and
$\sigma_{N}/\sigma_D$ are about 0.9~\%
below the cross section ratio reported by NMC.
A similar difference is observed in comparison to the SLAC data which
cover a smaller $x$ but the same $Q^2$ range than the HERMES data.
As the normalisation uncertainty of the present data is considerably
larger than that of the NMC data (0.4~\%), the $\sigma_{He}/\sigma_D$ and
$\sigma_{N}/\sigma_D$ results have been renormalised by 0.9~\%.
No such renormalisation has been applied to the Kr/D
cross section ratios.
For $x$ values below $x=0.1$, the present data on N/D and Kr/D are
slightly below the NMC
data but consistent within the present statistical and systematic uncertainties. 
\vspace*{-0.2cm}
 
\begin{figure} [H]
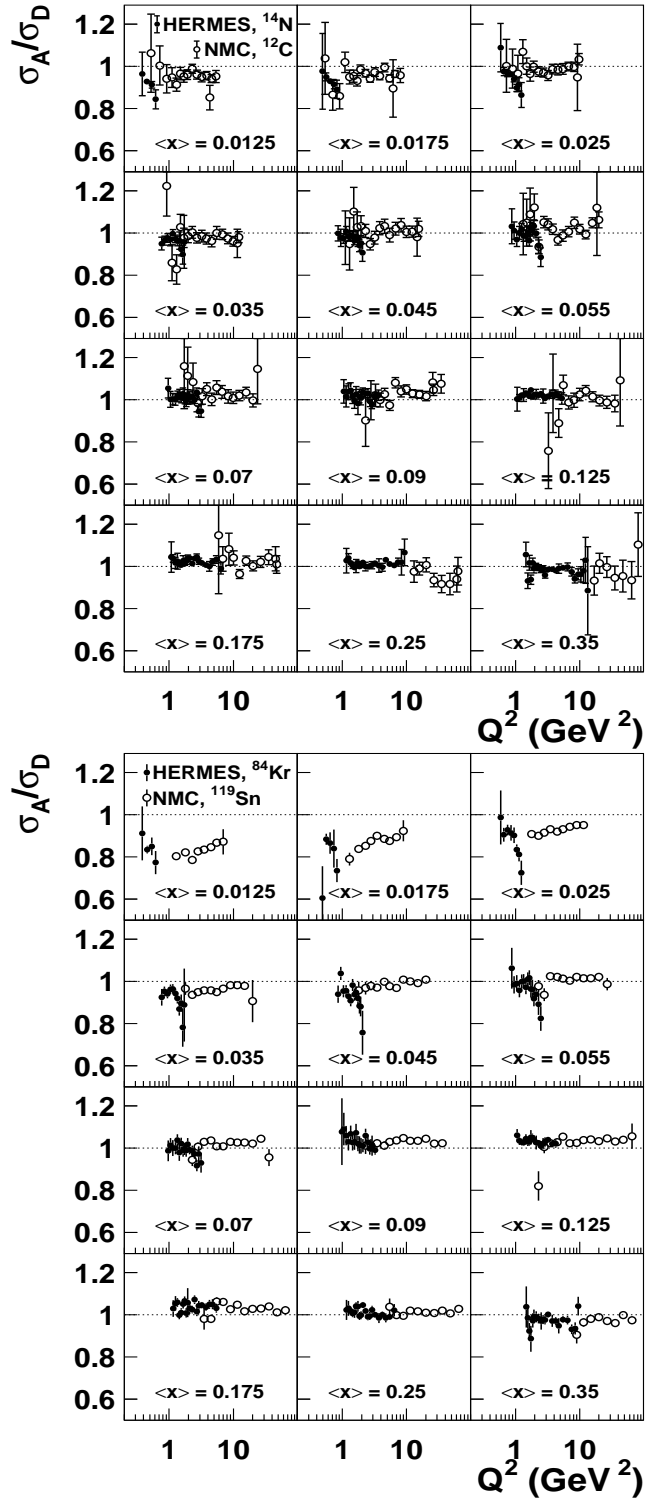

\begin{center}
\includegraphics[width=0.47\textwidth,height=0.42\textheight]{plots/xqbins_n2.epsi} 
\epsfxsize=\hsize 
\includegraphics[width=0.47\textwidth,height=0.42\textheight]{plots/xqbins_kr.epsi}
\epsfxsize=\hsize
\end{center}
\caption{
Ratio of isoscalar Born cross sections of
inclusive deep-inelastic lepton scattering from nucleus
$A$ and $D$ as function of $Q^2$ for fixed values of $x$.
The error bars represent the statistical
uncertainties.
The HERMES $^{14}$N/D data have been renormalised by
        0.9~\%.
}
\label{fig:qdep}
\end{figure}

The small discrepancy between the HERMES Kr/D and the NMC Sn/D data at 
$0.02 < x < 0.08$ is likely due to 
the positive $Q^2$ dependence observed in the NMC 
Sn/D data.   
This is better illustrated in Fig.~\ref{fig:qdep} 
where the $\sigma_{N}/\sigma_D$ and $\sigma_{Kr}/\sigma_D$ data are displayed 
as function of $Q^2$ for fixed values
of $x$ together with the NMC data on $\sigma_{C}/\sigma_D$ and 
$\sigma_{Sn}/\sigma_D$ respectively. 
For $x$ values below $x=0.08$, 
the NMC data on $\sigma_{Sn}/\sigma_D$  
show a significant positive $Q^2$ dependence resulting into a difference 
between the cross section ratio at the average $Q^2$ of the NMC data 
(as displayed in Fig.~\ref{fig:xdep}) and at the average $Q^2$ of the 
HERMES data. 
While no significant $Q^2$ dependence is observed in the cross section ratio of $^{14}$N/D, the large uncertainties in the HERMES Kr/D data do not allow to 
distinguish between a flat $Q^2$ dependence and an extrapolation of the 
small positive $Q^2$ dependence observed in the NMC measurement of 
$\sigma_{Sn}/\sigma_{D}$.

\begin{figure} [htb]
\begin{center}
\includegraphics[width=0.47\textwidth]{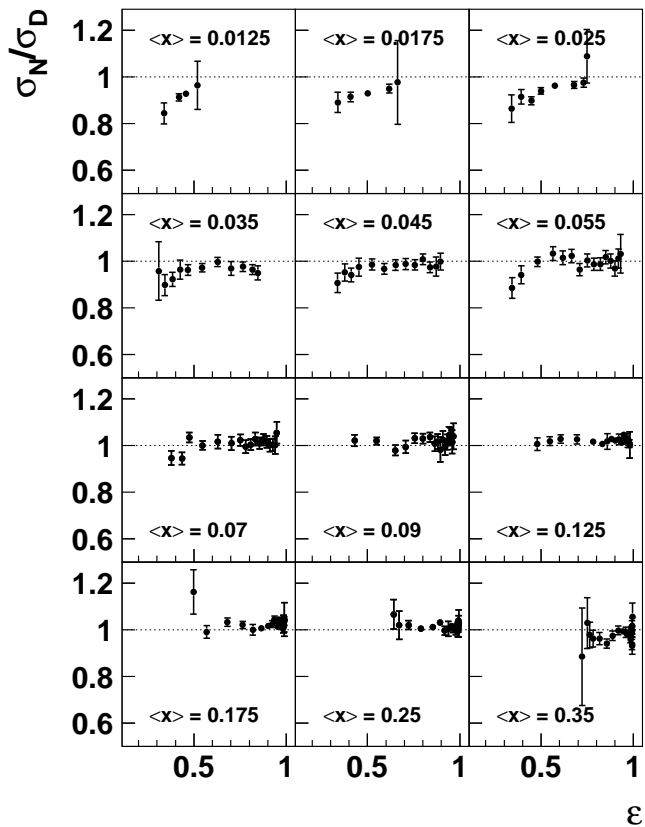}
\epsfxsize=\hsize
\end{center}
\caption{
Ratio of isoscalar Born cross sections of
inclusive deep-inelastic lepton scattering from nitrogen and deuterium
(renormalised by 0.9~\%)
as function of $\epsilon$ for fixed values of $x$.
The error bars represent the statistical
uncertainties.
}
\label{fig:edep}
\end{figure}

To investigate a possible A-dependence of $R(x,Q^2)$, the cross section ratios 
have been fitted as a function of 
$\epsilon$ for fixed values of $x$ (Eq.~3).
In these fits a parameterisation of
$R_D$ \cite{R1990} has been used, while the ratios
$R_A / R_D$ and $F_2^A / F_2^D$ have been treated as
free parameters.
A single value of $R_A / R_D$ and $F_2^A / F_2^D$ has been extracted
from each $x$-bin.
In this procedure it is assumed that
both $R_A / R_D$ and $F_2^A / F_2^D$ are
constant over the limited $Q^2$ range covered by the data in each $x$-bin.
The $\epsilon$-dependence of the $^{14}$N/D cross section ratio 
is shown in Fig.~\ref{fig:edep}. 
No sigificant $\epsilon$-dependence is observed.
Similar conclusions hold for the other target nuclei. 

\begin{figure} [htb]
\begin{center}
\includegraphics[width=0.47\textwidth]{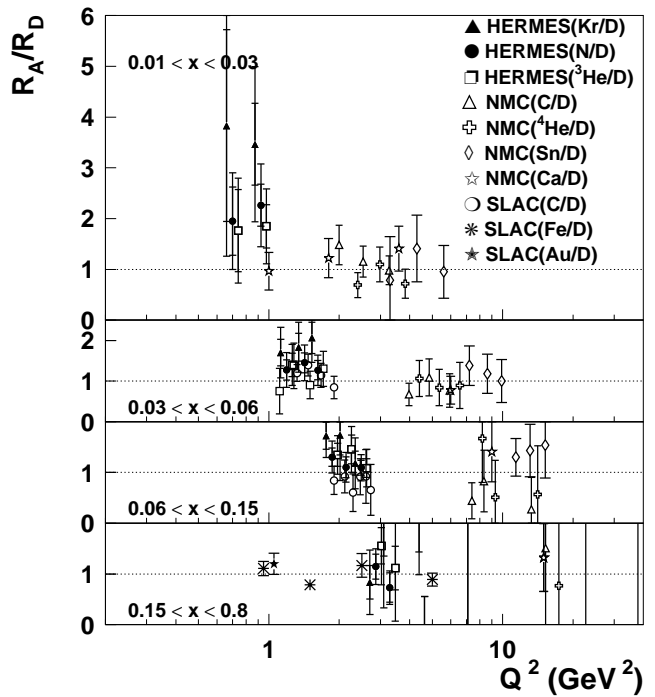}
\caption[]{\label{fig:rq-all}
The isoscalar-corrected ratio $R_A / R_D$ for several nuclei (A) with respect to 
deuterium as a function of $Q^2$ for four different
$x$ bins. 
The open triangles ($^{12}$C) and crosses ($^{4}$He)
have been derived from the NMC data using Eq.~3. The other
SLAC \cite{AuD} and NMC data \cite{SnC} displayed have been derived from
published values of $\Delta R = R_A - R_D$ and a parameterisation
\cite{R1990} for $R_D$. The inner error bars represent the statistical
uncertainty and include the correlated error in $F_2^A/F_2^D$. The outer
error bars represent the quadratic sum of the statistical and systematic uncertainties.
In the upper panel the HERMES results at the lowest $Q^2$ value 
have been suppressed because of its large error bar.
}
\end{center}
\end{figure}

The values of $F_2^A/F_2^D$ derived from the fit of the HERMES data
are found to be consistent with previous measurements of NMC and SLAC. 
The resulting values of $R_A / R_D$ are shown in Fig.~\ref{fig:rq-all}.
Also shown are 
the results of previous studies of the $A$-dependence of $R$.
Existing data are usually represented in terms
of $\Delta R = R_A - R_D$.
The published values of $\Delta R$ \cite{SLAC,AuD,SnC}
have been converted to $R_A /R_D$
using a parameterisation for $R_D$ \cite{R1990}, 
and added to Fig.~\ref{fig:rq-all}.
All data are found to be 
consistent with unity.

\noindent

\begin{figure} [htb]
\begin{center}
\includegraphics[width=0.47\textwidth]{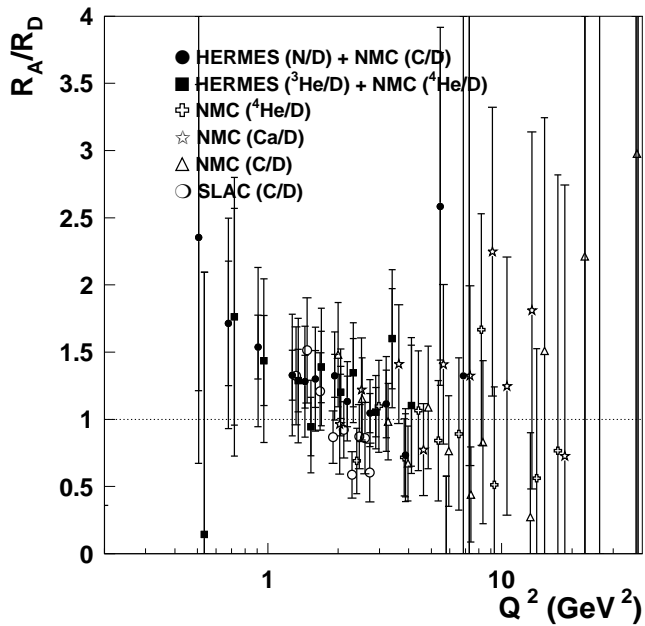}
\caption[]{\label{fig:rfix}
The isoscalar-corrected ratio $R_A / R_D$ for several nuclei (A) with respect to
deuterium as a function of $Q^2$. The HERMES and NMC data have been combined 
in the determination of $R_A / R_D$.
The other data are the same as in Fig.~\ref{fig:rq-all}.
}
\end{center}
\end{figure}

At low $x$, the 
HERMES cross section ratios on $^3$He and $^{14}$N and the NMC    
measurements on $^4$He and $^{12}$C have some common $Q^2$ range. 
While the NMC measurements at these $x$ and $Q^2$ values have $\epsilon$ 
values close to unity, the HERMES data cover a typical $\epsilon$ range 
of $0.4 < \epsilon  < 0.7$.
Combining the two measurements thus increases the precision on $R_A/R_D$.
Because of the different experimental conditions of the NMC 
measurements on the heavy targets no such overlap in $x$ and $Q^2$ 
exists for the HERMES krypton data and the NMC data on iron or tin. 
Therefore the determination of $R_A/R_D$ on heavy nuclei is not 
improved by combining the HERMES and NMC data.  
The results of the fits to the HERMES and NMC data on helium and 
nitrogen (carbon) 
are displayed in Fig.~\ref{fig:rfix} as 
a function of $Q^2$ 
together with all other 
measurements of $R_A/R_D$ on light and medium heavy nuclei. 
For $Q^2$ values between 0.5 and 20 GeV$^2$ 
and nuclei from He to Ca, $R_A$ is found to be consistent with 
the $R$ parametrisation of Whitlow et al~\cite{R1990}. 
As throughout this paper, this $R$ parametrisation 
has been choosen in this comparison  
because it is dominated by data on the proton and the deuteron.
In contrast,  
the more recent parametrisation by Abe et al.~\cite{SLAC} 
is significantly influenced by nuclear data. 
The influence of the choice in the $R$ parametrisation is however very small. 
Averaging over all measurements of $R_A/R_D$ for light and medium 
heavy nuclei gives an average value for $R_A/R_D$ of $0.99 \pm 0.03$, 
a result which is unchanged if the data on the heavier nuclei are included in 
the average.  

%

\noindent

In summary, deep-inelastic positron scattering data on $^2$H,
$^3$He, $^{14}$N and $^{84}$Kr are presented. 
The results extracted for the ratios of the DIS cross sections on nuclei 
to those on the corresponding sets of free nucleons is in 
good agreement with      
the results from previous measurements, after the 
data were corrected for a previously unrecognised A-dependent 
tracking inefficiency. No significant $Q^2$ dependence 
is observed over the wide range in $Q^2$ covered by the combined 
data set of HERMES and NMC. 
Values for the ratio of
$R_A / R_D$ with $R$ the ratio $\sigma_L / \sigma_T$ of longitudinal
to transverse DIS cross sections have been derived
from the dependence of the data on the virtual photon
polarisation parameter $\epsilon$ and found to be consistent with unity.
The data presented in this paper extend the kinematic range in which 
$R_A$ is found to be equal to $R_D$ down to $x=0.01$ and $Q^2=0.5$ GeV$^2$.

We gratefully acknowledge the DESY management for its support and
the DESY staff and the staffs of the collaborating institutions.
This work was supported by
the FWO-Flanders, Belgium;
the Natural Sciences and Engineering Research Council of Canada;
the INTAS contribution from  the European Commission;
the European Commission IHP program under contract HPRN-CT-2000-00130;
the German Bundesministerium f\"ur Bildung und Forschung (BMBF); 
the Deutscher Akademischer Austauschdienst (DAAD);
the Italian Istituto Nazionale di Fisica Nucleare (INFN);
Monbusho International Scientific Research Program, JSPS, and Toray
Science Foundation of Japan;
the Dutch Foundation for Fundamenteel Onderzoek der Materie (FOM);
the U.K. Particle Physics and Astronomy Research Council; and
the U.S. Department of Energy and National Science Foundation.


\end{document}